\documentstyle[12pt,epsfig]{article}

\newcommand{\bee}   {\begin{equation}}
\newcommand{\ene}   {\end{equation}}
\newcommand{\beqa}  {\begin{eqnarray}}
\newcommand{\enqa}  {\end{eqnarray}}
\newcommand{\bea}   {\begin{array}}
\newcommand{\ena}   {\end{array}}

\def\ru1{\rule[-0.4truecm]{0mm}{1truecm}}

\def\ga{\gamma}
\def\als{{\alpha_s}}
\def\upa{\uparrow}
\def\alp{\frac{\als(Q^2)}{\pi}}
\def\ald{3.58}
\def\alt{20.22}

\def\dqtns{\Delta\tilde q_{NS}}
\def\xdqns{x\Delta q_{NS}}
\def\dPt{\Delta\tilde P_{qq}}

\begin{document}

$~$
\hfill{ \mbox{Napoli Preprint {\bf DSF-T-96/43}}}

\hfill{ \mbox{ hep-ph/9704270 } }

\vspace{2truecm}

\centerline{
{\LARGE Low $x$ Behaviour of the Isovector Nucleon Polarized}}

\vspace{0.5truecm}

\centerline{
{\LARGE Structure Function and the Bjorken Sum Rule}}

\vspace{1.8truecm}

\centerline{
{\large Franco BUCCELLA, Ofelia PISANTI, Pietro SANTORELLI}}

\vspace{0.1truecm}

\begin{center}
{\it
Dipartimento di Scienze Fisiche, Universit\`a ``Federico II'',\\
Pad. 19 Mostra d'Oltremare, 00195 Napoli, Italy \\
INFN, Sezione di Napoli,\\
Pad. 20 Mostra d'Oltremare, 00195 Napoli, Italy}
\end{center}

\vspace{1.2truecm}

\begin{abstract}
The combination $g_1^p(x) - g_1^n(x)$ is derived from SLAC data on
polarized proton and deuteron targets, evaluated at $Q^2\,=\,10\,GeV^2$,
and compared with the results of SMC experiment. The agreement is
satisfactory except for the points at the three lowest $x$, which have an
important role in the SMC evaluation of the l.h.s. of the Bjorken sum
rule.
\end{abstract}

\newpage
\baselineskip   = 18pt

The measurement by EMC \cite{emc} of
\bee
I_p = \int_0^1 g_1^p(x) dx = 0.126\,\pm\,0.010\,\pm\,0.015\,,
\ene
smaller than the prediction of the Ellis and Jaffe (EJ) sum rule
\cite{elljaf},
$F/2 - D/18\;(F\,=\,0.46\,\pm\,0.04,~D\,=\,0.79\,\pm\,0.04$
\cite{FD}), stimulated experimental and theoretical work on the spin
structure of the nucleon.

The data on $g_1^p(x)$ were in fair agreement with a previous calculation
by the Bari group \cite{Bari} in the framework of a model where the
Bjorken (Bj) sum rule \cite{bjork} is not obeyed. To explain the low value
of $I_p$ it was assumed \cite{deltas} that the strange quark carries a
large negative component of the proton (and neutron) spin. However, a
phenomenological study \cite{PrepSoff} of the strange sea (from charm
production) shows that its non-diffractive part is very small and, for
positivity reasons, cannot account for the polarization of the strange sea
needed to explain the EMC result.

The defect in the EJ sum rule for the proton is in part explained by the
QCD corrections \cite{Kodaira}, and what is left may be accounted by a
large positive projection \cite{anom} ($\sim 2$) of the gluon spin along
the proton spin (to be compensated by a negative contribution of the
orbital angular momentum). In this framework the Bj sum rule, apart from
$Q^2$-dependent QCD corrections \cite{Kodaira}, should be valid:
\bee
I_p - I_n = \frac{1}{6} \frac{G_A}{G_V} \left[ 1 - \alp - \ald \left( \alp
\right)^2 - \alt \left( \alp \right)^3 \right].
\ene

A different explanation \cite{BuccSoff}, which relates the defects in the
EJ sum rule to the defect in the Gottfried sum rule \cite{gottfr}, implies
a violation of the Bj sum rule of ($-0.025\,\pm\,0.007$).

Experimentally, one has measurements on $g_1^p(x)$ and $g_1^d(x)$ at SLAC
\cite{slac-p-d} (at $Q^2\,=\,3\,GeV^2$) and CERN \cite{smc-p-d} (at
$Q^2\,=\,10\,GeV^2$): within their experimental errors they do not show
evidence for a violation of the Bj sum rule. A more severe test is coming
from the forthcoming measurements \cite{Hughes} with $He^3$ polarized
targets, which practically give directly $g_1^n(x)$; the experiment already
performed with this target at $Q^2\,=\,2\,GeV^2$ \cite{e142} shows the good
precision one can reach. It is worth stressing that the errors in
measuring $g_1^p(x)$ and $g_1^n(x)$ count twice less in testing the Bj sum
rule with respect to measurements on $g_1^p(x)$ and $g_1^d(x)$.

It is also worth observing that an important contribution to the l.h.s of
the Bj sum rule at $Q^2\,=\,10\,GeV^2$ comes from the low $x$ region,
where the lowest point for $xg_1^p(x)$ is higher than the subsequent
points and $xg_1^d(x)$ becomes negative. This behaviour is not reproduced
with simple parameterizations of the polarized structure functions in
terms of the contributions of the valence quarks and of the gluons, either
by taking as input the SMC data \cite{gerst} or by getting predictions at
$Q^2\,=\,10\,GeV^2$ from the evolution of the result of a fit of the E143
data at $Q^2\,=\,3\,GeV^2$ \cite{BBPSS}.

Here we want to focus our attention on the isovector part of $g_1^N(x)$,
the one relevant for the Bj sum rule and not influenced in its evolution
by the isoscalar terms $\Delta G(x)$ and $\Delta s(x) + \Delta\bar s(x)$.
Our strategy will be to get $g_1^p(x)\,-\,g_1^n(x)$ from SLAC data and to
compare the evolved structure function with CERN data.

To construct the isovector combination $g_1^p(x)\,-\,g_1^n(x)$ from SLAC
data we have the technical problem that only the first lowest fourteen
values of $x$ coincide in the two experiments which measure $g_1^p(x)$ and
$g_1^d(x)$, while at higher $x$ one has other fourteen values of $x$ for
the proton and seven different values for the deuteron. To get a reliable
interpolation of proton data in correspondence of the seven highest values
of $x$ measured for the deuteron, we take the following parameterization
for $g_1^p(x)$,
\bee
g_1^p(x)\,=\,A_1\,x^{A_2}\,(1-x)^{A_3}\exp\{A_4\,x\}\;,
\ene
and find, from the fit to the values of $g_1^p(x)$,
\bee
\begin{array}{lcl}
A_1\,=\,0.37\,\pm\,0.01\,, &~~~& A_2\,=\,0.046\,\pm\,0.013\,, \\
A_3\,=\,1.32\,\pm\,0.07\,,   &~~~& A_4\,=\,-0.41\,\pm\,0.09\;. 
\end{array}
\ene

To get $g_1^p(x) - g_1^n(x)$ from $g_1^p(x)$ and $g_1^d(x)$, we keep into
account the small D-wave component in the deuteron ground state, which
implies, with $\omega_D\,=\,0.058$,
\bee
g_1^d(x) \,=\, \left( 1-\frac{3}{2}\,\omega_D \right)\, \frac{g_1^p(x) +
g_1^n(x)}{2}\;.
\ene

We write $g_1^p(x) - g_1^n(x)$ at $Q_0^2 \,= \, 3\, GeV^2$ as the sum of
the contributions of the valence quarks and of a possible term coming from
the sea, $\Delta {\cal S}(x)$ (indeed the sea is responsible of the defect
in the Gottfried sum rule),
\bee
g_1^p(x) - g_1^n(x) = \frac{1}{6}\, \left(\Delta u(x) - \Delta d(x) +
\Delta {\cal S}(x)\right)\;,
\ene
with the standard parameterization,
\beqa
x\Delta u(x) & = & \eta_u \, A_u \, x^{\alpha}\, (1-x)^{\beta_u}\,
(1+\ga\,x)\;, \nonumber \\
x\Delta d(x) & = & \eta_d \, A_d \, x^{\alpha}\,(1-x)^{\beta_d}\,
(1+\ga\,x)\;.
\enqa
Since in the quark parton model $\Delta {\cal S}$ should correspond to
$\Delta\bar u - \Delta\bar d$, we adopt for it the same shape of the
unpolarized sea, $\bar u - \bar d$ \cite{mrs}, which gives
\bee
x\Delta {\cal S}(x) = \eta_{{\cal S}}\,A_{{\cal S}}\, x^{0.3}\,
(1-x)^{10.1}\, (1+64.9\,x)\;.
\ene
The normalization factors $A_q$ and $A_{{\cal S}}$ are defined in such a
way that the first moments are:
\bee
\bea{lcl}
\Delta q        & = & \eta_q\;, \\
\Delta {\cal S} & = & \eta_{{\cal S}}\;.
\ena
\ene

To remove the ambiguity coming from describing one function as a
combination of two or three functions, we require that, for $x\, \geq\,
0.2$,
\bee
g_1^p(x) \,=\, \frac{2}{9}\,\Delta u(x) + \frac{1}{18} \Delta d(x)\;.
\ene
This constraint finds its motivations in the fact that the region
considered is dominated by the valence partons. Finally, to keep into
account that the $u^{\upa}$ quark is dominating at high $x$
\cite{Betal,BMT}, to get $\beta_d$ larger than $\beta_u$ we require
\bee
\beta_d\,\geq\,2.5\;.
\ene 

We have considered several options for $\eta_u$, $\eta_d$ and $\eta_{{\cal
S}}$, but, since the main conclusion is the same, we present in Table
\ref{t:ris} the results for three particular choices, which are good
examples of what happens also for the other options. For the three options
we fix $\eta_d$ to its QCD value, with the corrections up to the third
order in $\als$,
\bee
\eta_d \,=\, -0.26\pm 0.02\;.
\ene

For the first option ({\bf 1}) we also fix $\eta_u$ to its QCD value,
again up to ${\cal O}(\als^3)$,
\bee
\eta_u \,=\, 0.76\pm 0.04\;.
\ene

Instead, for the second ({\bf 2}) and the third ({\bf 3}) option $\eta_u$
is left free and for the second one $\eta_{{\cal S}}$ is fixed in such a
way that the Bj sum rule is obeyed. In conclusion, only in the third
option we leave open the possibility that the Bj sum rule is violated.

The comparison with the data on $x(g_1^p(x) - g_1^n(x))$ and $xg_1^p(x)$
(for $x\,\geq\,0.2$), and the distributions $x\,\Delta u(x)$, $x\,\Delta
d(x)$ and $x\,\Delta {\cal S}(x)$ for the three fits are reported in
Figures 1, 2 and 3 respectively. The fits do not differ so much and have
satisfactory $\chi^2$. \\
The third option supplies a favourable test to the Bj sum rule with a
defect of only $2\%$ well consistent with the experimental errors.\\
The intercept for the first option, $\alpha = 0.63 \,\pm\,0.12$, is
slightly less than the value $0.77 \,\pm\,0.01$ mostly dependent on the
non-singlet (NS) unpolarized structure function, $F_2^p(x) - F_2^n(x)$ and
$xF_3(x)$, obtained in Ref. \cite{BMT}; the agreement is better for the
other two options. The difference $\beta_d - \beta_u$ is around 1 and
$\ga$ is consistent with zero.

To get the polarized NS distribution at $Q^2\,=\,10\,GeV^2$, one has the
Dokshitzer-Gribov-Lipatov-Altarelli-Parisi (DGLAP) equations \cite{dglap}
in the variable $t \equiv \ln Q^2/\Lambda^2_{QCD}$ ($\dqtns\equiv \xdqns$):
\bee
\frac{d}{dt}\dqtns(x,t) = \frac{\als (t)}{2\pi} \int_x^1 \frac{dz}{z} \dPt
(z) \dqtns \left( \frac{x}{z},t \right)\;,
\label{e:APpolns}
\ene
with, at next to leading order (NLO) in $\als$ \cite{polspl},
\bee
\dPt (z) = \dPt^{(0)} + \frac{\als}{2\,\pi} \dPt^{(1)}\,,
\ene
where $\dPt^{(0)}$ is the NS polarized splitting function at leading order
\cite{dglap} and $\dPt^{(1)}$ is the NLO one \cite{npolspl}. We fix
$\Lambda_{QCD}^{(4)} = 359\, MeV$ to get at NLO, with $n_f=4$,
$\als(3\,GeV^2) = 0.35 \pm 0.05$.

To solve the DGLAP equations, we use a method which, by expanding $\dqtns$
into a truncated series of Chebyshev polynomials \cite{kwiez},  gives rise
to a system of coupled differential equations for the values of $\dqtns$
in the points corresponding to the nodes of the Chebyshev polynomials.

With the initial conditions, given by the fits to SLAC data reported in
Table \ref{t:ris}, we get the corresponding evolved distributions at
$Q^2\,=\,10\,GeV^2$, which are compared in Figure 4 with the values of
$x\,(g_1^p(x)-g_1^n(x))$ deduced by the SMC measurements on $g_1^p(x)$ and
$g_1^d(x)$ (for the deuteron, instead of the value of Ref. \cite{smc-p-d},
we take the more recent ones presented at Morillon and reported to us in
Ref. \cite{Hughes}).

In Table \ref{t:ris} we report also the corresponding $\chi^2$ for the
comparison between the evoluted distribution and SMC data. The large
experimental errors help to get $\chi^2_{ISOVEC}$ in the range (11,13),
but for all the three options, as well as for the ones we did not report
in Table \ref{t:ris}, the curves obtained from the evolution of SLAC data
pass below the points corresponding to the three lowest values of $x$,
implying a behaviour in the limit $x\rightarrow 0$ different from what
could be derived from SMC data. Since the low $x$ region gives an
important contribution to the l.h.s. of the Bj sum rule evaluated by SMC
Collaboration, the fact that, even allowing for isovector polarization in
the sea, one is not able to recover the low $x$ behaviour suggested by the
three points at the lowest $x$, does not allow to be confident on
conclusions about the validity of the Bj sum rule based on these three
points in a range of $x$ not explored by SLAC.


%

\newpage

\begin{table}[ht]
\begin{center}
TABLE I\\
\vspace{.6truecm}
\begin{small}
\begin{tabular}{|c|c|c|c|} 
\hline \ru1
 & {\bf 1} & {\bf 2} & {\bf 3}	\\
\hline\hline\ru1
$\alpha$ & $0.63\pm 0.12$ & $0.71\pm 0.19$ & $0.70\pm 0.23$ \\
\hline\ru1
$\beta_u$ & $1.5\pm 0.4$ & $1.6\pm 0.4$ & $1.4\pm 0.4$ \\
\hline\ru1
$\beta_d$  & $2.5\pm 0.9$ & $2.5\pm 1.3$ & $2.5\pm 1.3$ \\
\hline\ru1
$\ga$ & $1.6\pm 2.1$ & $1.9\pm 2.7$ & $1.1\pm 2.3$ \\
\hline\ru1
$\eta_u$ & $0.76\pm 0.04$ & $0.70\pm 0.08$ & $0.74\pm 0.07$ \\
\hline\ru1
$\eta_d$ & $-0.26\pm 0.02$ & $-0.26\pm 0.02$ & $-0.26\pm 0.02$ \\
\hline\ru1
$\eta_{{\cal S}}$  & $0$ & $0.06\pm 0.07$ & $0$ \\
\hline\ru1
$\chi^2/dof$ & $32.06/29\; = \;1.11$ & $31.31/28 \; = \; 1.12$ & $31.96/28
\; = \; 1.14$ \\
\hline\ru1
$\chi^2_{ISOVEC}$ & $11.7$ & $12.7$ & $12.7$ \\
\hline
\end{tabular}
\end{small}
\end{center}
\caption{The results of the options {\bf 1}, {\bf 2} and {\bf 3} for the
values of the parameters of the fits at $Q^2 = 3\,GeV^2$. In the last row
we show the $\chi^2$ for the comparison of the three options, evoluted to
$Q^2 = 10\,GeV^2$, with the SMC data (see text).}
\label{t:ris}
\end{table}

\newpage

\section*{Figure Captions}

\begin{itemize}
\item[Fig. 1]
The results corresponding to fits {\bf 1} (solid line), {\bf 2} (dashed
line), and {\bf 3} (dotted line) are compared with the SLAC data on
$x(g^p_1(x)-g^n_1(x))$ at $<\!\!Q^2\!\!>= 3\,GeV^2$ from ref.
\cite{slac-p-d}.

\item[Fig. 2]
Same as Fig. 1 for the proton SLAC data for $xg^p_1(x)$ from ref.
\cite{slac-p-d}. Note that only the experimental data used for the fits
are plotted.

\item[Fig. 3]
The distributions $x\Delta u$, $x\Delta d$, and $x\Delta {\cal S}$ are
plotted for fit {\bf 1} (solid line), fit {\bf 2} (dashed line) and fit
{\bf 3} (dotted line). Note that for fit {\bf 1} and {\bf 3} $x\Delta
{\cal S}$ is zero.

\item[Fig. 4]
The evolution to $Q^2 = 10\,GeV^2$ of the results of fits {\bf 1} (solid
line), {\bf 2} (dashed line), and {\bf 3} (dotted line) are compared with
the SMC data on $x(g^p_1(x)-g^n_1(x))$ at $<\!\!Q^2\!\!>= 10\,GeV^2$ from
Ref. \cite{smc-p-d} (the deuteron data are from Ref. \cite{Hughes}).

\end{itemize}

\newpage
\pagestyle{empty}

\begin{figure}[p]
\centerline{\epsfig{file=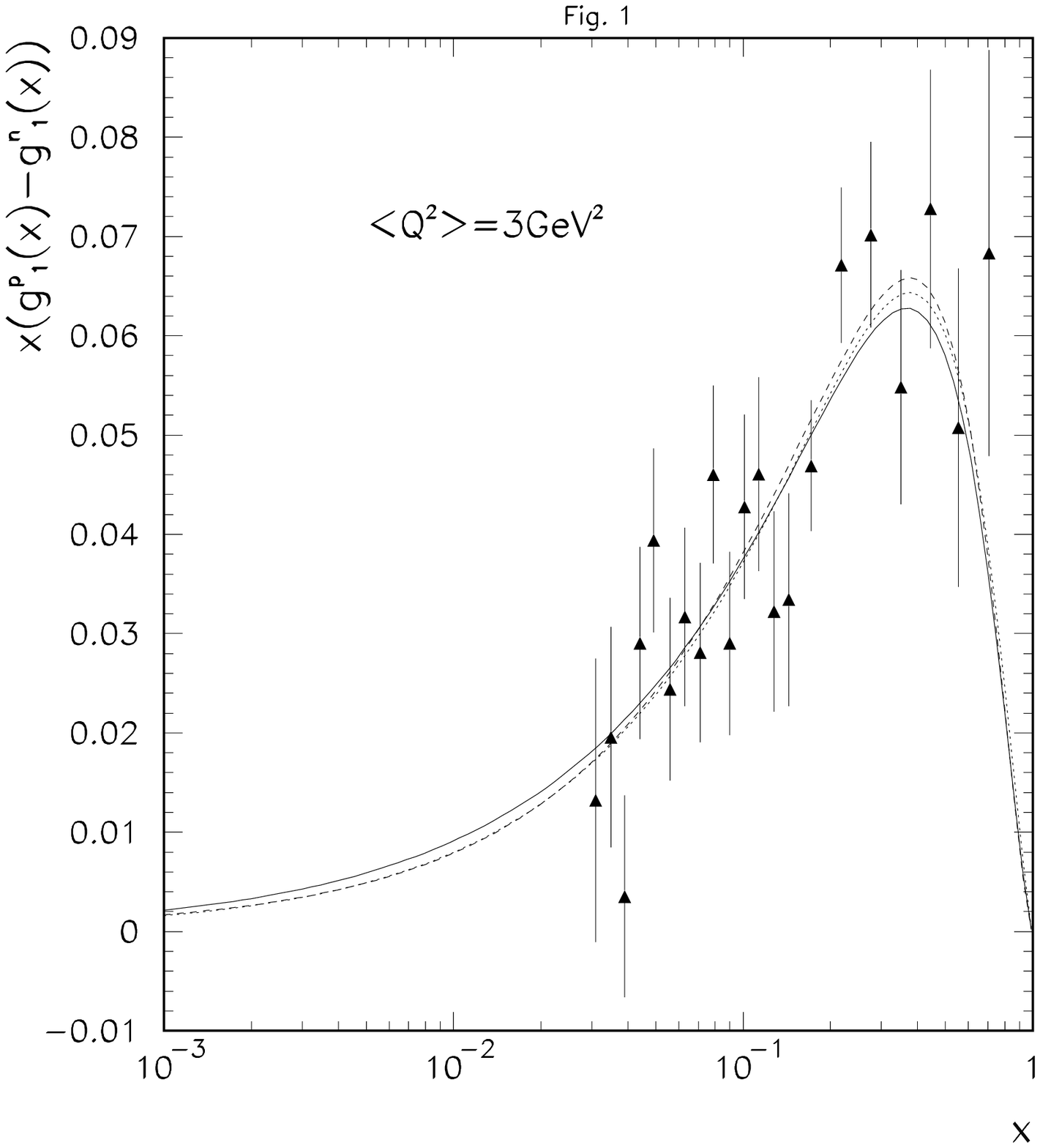,height=15cm}}
\label{fig:fig1}
\end{figure}

\newpage

\begin{figure}[p]
\centerline{\epsfig{file=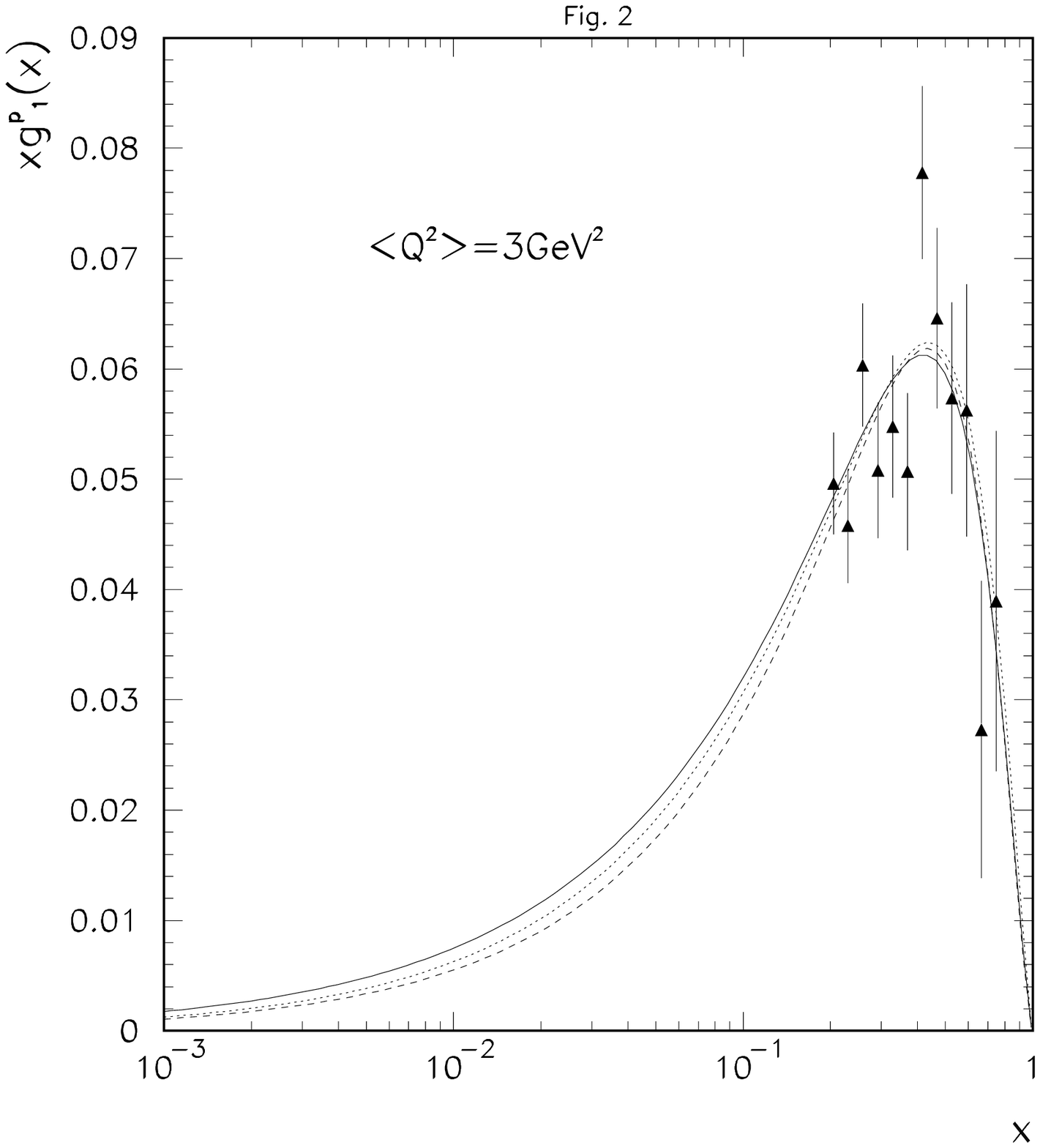,height=15cm}}
\label{fig:fig2}
\end{figure}

\newpage

\begin{figure}[p]
\centerline{\epsfig{file=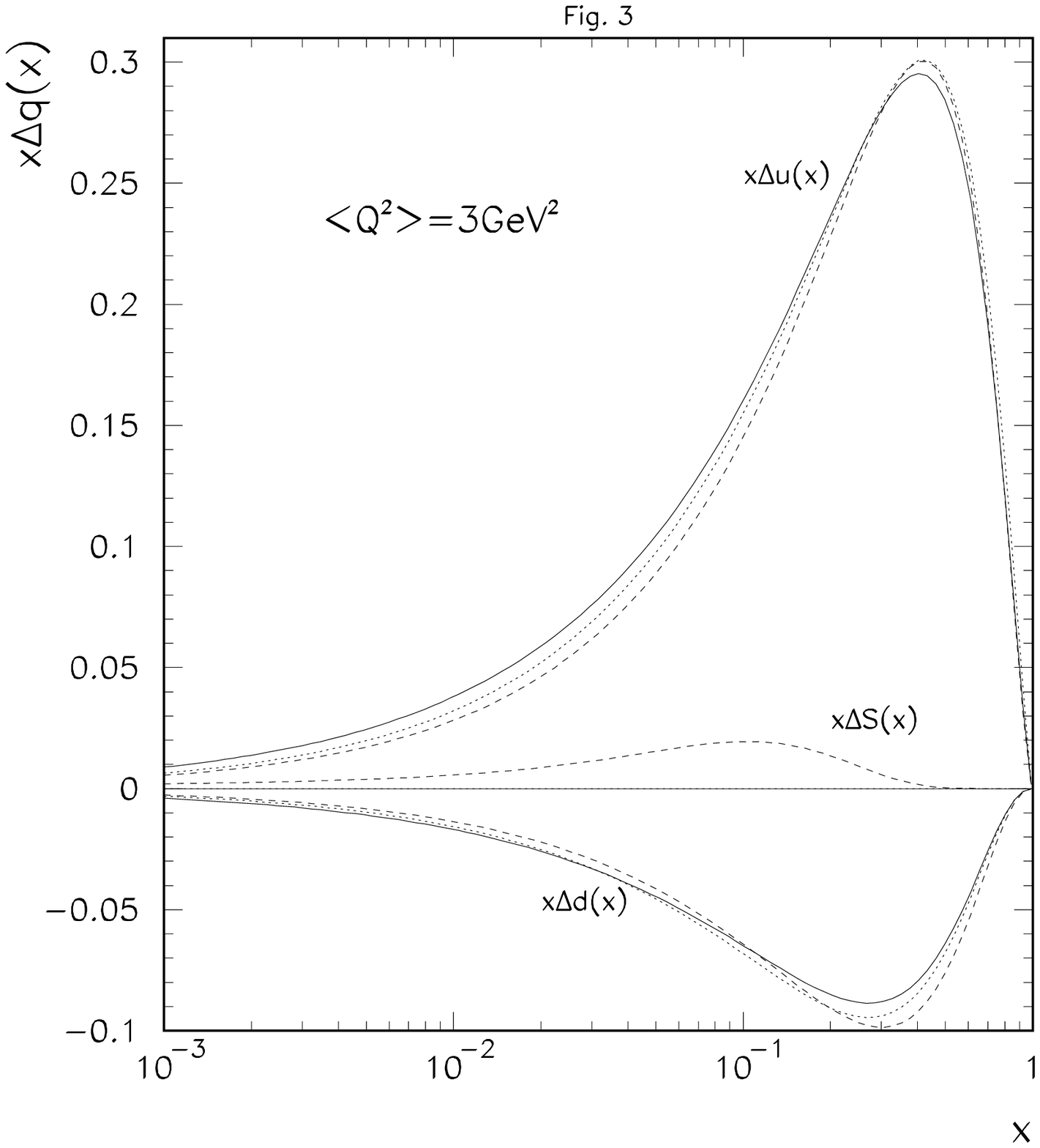,height=15cm}}
\label{fig:fig3}
\end{figure}

\newpage

\begin{figure}[p]
\centerline{\epsfig{file=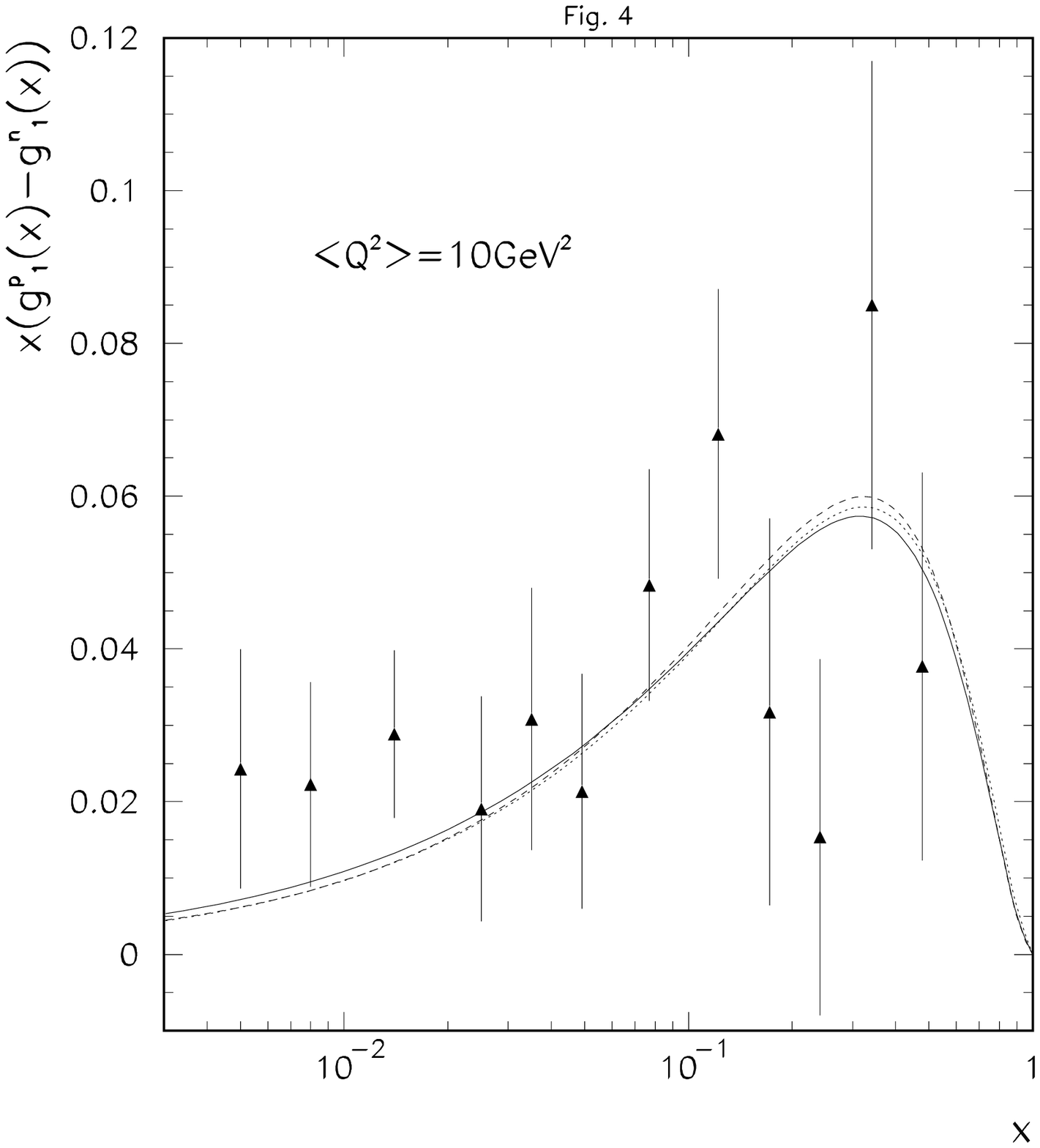,height=15cm}}
\label{fig:fig4}
\end{figure}

\end{document}